\documentclass[preprint]{emulateapj}

\newcommand{\Swift}{{\it Swift}}

\def\simlt{\mathrel{\hbox{\rlap{\hbox{\lower4pt\hbox{$\sim$}}}\hbox{$<$}}}}
\def\simgt{\mathrel{\hbox{\rlap{\hbox{\lower4pt\hbox{$\sim$}}}\hbox{$>$}}}}
\def\arcdeg{\hbox{$^\circ$}}

\newcommand{\peryear}{yr$^{-1}$}

\begin{document}
\title{The Local Rate and the Progenitor Lifetimes of Short-Hard Gamma-Ray Bursts:
       Synthesis and Predictions for LIGO}

\author{Ehud Nakar, Avishay Gal-Yam\altaffilmark{1}}

\affil{Division of Physics, Mathematics and Astronomy, The
California Institute of Technology, Pasadena, CA, 91125, USA}
\email{udini@tapir.caltech.edu}

\and

\author{Derek B. Fox}

\affil{Department of Astronomy \& Astrophysics, 525 Davey
Laboratory, The Pennsylvania State University, University Park, PA
16802, USA}

\altaffiltext{1}{Hubble Fellow}

%\maketitle

\begin{abstract}

The recent discovery of the first four afterglows of short-hard
gamma-ray bursts (SHBs), and the properties inferred for these
bursts -- as well as for  four other SHBs with known or constrained
redshift -- suggests that these events typically result from
long-lived progenitor systems. The most popular model invokes the
merger of two compact objects, either double neutron star (DNS)
binaries or neutron star-black hole (NS-BH) systems. Such events
should emit a significant fraction of their ultimate binding energy
in gravitational radiation in the frequency range that is accessible
to current and next-generation ground-based gravitational-wave
observatories. In this work we combine the census of SHB
observations with refined theoretical analysis to perform a critical
evaluation of the compact binary model. We then explore the
implications for gravitational wave detection of these events.
Beginning from the measured star-formation rate through cosmic time,
we consider what intrinsic luminosity and lifetime distributions can
reproduce both the known SHB redshifts and luminosities as well as
the peak flux distribution of the large BATSE SHB sample. We find
that: (1) The {\it typical} progenitor lifetime, $\tau_*$, is long.
Assuming lognormal lifetime distribution $\tau_*>4[1]$ Gyr (at
$2[3]\sigma$ c.l.). If the lifetime distribution is a power-law with
index $\eta$ then $\eta>-0.5 [-1]$ (at $2[3]\sigma$ c.l.). This
result is difficult to reconcile with the properties of the observed
galactic DNS population, suggesting that if SHBs do result from DNS
mergers then the observed galactic binaries do not represent the
cosmic DNS population. (2) We find that the local rate of SHBs is at
least ${\cal R}_{SHB} \gtrsim 10 ~\rm Gpc^{-3~} yr^{-1}$ and may be
higher by several orders of magnitude, significantly above previous
estimates. (3) We find that, assuming that SHBs do result from
compact binaries, our predictions for the LIGO and VIRGO event rates
are encouraging: The chance for detection by current facilities is
not negligible, while a coincident observation of electromagnetic
and gravitational radiation from an SHB is guaranteed for
next-generation observatories.

\end{abstract}

%%%%%%%%%%%%%%%%%%%%%%%%%%%%%%%%%%%%%%%%%%%%%%%%%%%%%%%%%%%%%%%%%%%%%%

\section{Introduction}

More than a decade ago it was realized that Gamma-Ray Bursts (GRBs)
can be divided into two well-defined sub-populations: the majority
of the observed bursts (about 3/4) had durations longer than
$\sim2$s and relatively softer observed spectra, and the minority
($\sim1/4$) had short durations and harder observed spectra
\citep{Kouveliotou93}. Accordingly, the two components of this
bimodal population are often referred to as long-soft GRBs and
short-hard GRBs (SHBs hereafter).

Observational evidence accumulated in the last few years has
conclusively shown that long-soft GRBs are associated with
supernovae (SNe), and result from the death of short-lived massive
stars \citep{galama98,Kulkarni98,bloom99,bloom02,
stanek03,hjorth03,matheson03,lipkin04,gal-yam04,malesani04,cobb04,thomsen04,
Soderberg05}. Comparable studies of SHBs were not conducted due to
observational difficulties in obtaining accurate localizations for
these events.

A breakthrough in the study of SHBs occurred earlier this year when
the {\it
Swift}\footnote{http://swift.gsfc.nasa.gov/docs/swift/swiftsc.html}
and {\it HETE-2}\footnote{http://space.mit.edu/HETE/Welcome.html}
satellites provided the first timely and accurate localizations of
SHBs 050509b, 050709 and 050724, leading to the detection of
afterglow emission in X-ray, optical, IR and radio wavelengths, as
well as to the identification and study of the host galaxies of
these events \citep{bloom05,kulkarni05,Gehrels05,
Castro05b,Prochaska05,Fox05,Hjorth05,Covino06,Berger05,hwf+05}.
These studies provide three strong indications that SHBs result from
a different type of physical progenitor system. First, in all three
events there are no indications of an associated SN, to within
strict limits. SNe similar to those associated with long-soft GRBs
would have been easily detected. Second, two of these events
occurred within early-type galaxies, with little or no recent star
formation. None of the many tens of long-soft GRBs with comparable
data are detected in such galaxies. Finally, all of these bursts
occurred relatively nearby, at redshifts $z<0.3$, in contrast with
long GRBs that are typically at $z>1$. Taken together these
observational facts strongly indicate that SHBs have a different
type of progenitor system, and that some of these progenitors are
long-lived ($\gtrsim1$ Gyr).

The most popular model (e.g., \citealt{Eichler89, Narayan92}) for
SHBs invokes the merger of two compact objects, such as double
neutron stars (DNS) or a neutron star and a black hole (NS-BH). Such models
are appealing because they predict events with comparable timescale and energy
release to those observed in short bursts. Such
considerations, combined with the recent results showing that SHBs
require a long-lived progenitor system, led several groups to
suggest that SHBs result from compact binary mergers (e.g.,
\citealt{bloom05,Fox05,Berger05}). More detailed comparisons with
model predictions were limited by the small numbers of SHBs with
known host galaxies and redshifts.

\citet{gal-yam05} have expanded the sample of SHBs available for
study using new and archival observations of ``historical'' SHBs,
allowing them to establish a statistically significant association
between two additional SHBs (790613 and 000607) and their probable
host galaxies or clusters. Combining these data with observations of
the recent {\it Swift} and {\it HETE-2} SHBs discussed above, and an
additional {\it Swift} burst (SHB 050813) possibly located in a
$z=0.72$ galaxy cluster \citep{Gladders05,berger05b,Prochaska05},
\citet{gal-yam05} confronted predictions from binary NS merger
models \citep{Ando04,Guetta05} with the properties of the observed
SHB sample. They found that the model predictions are difficult to
reconcile with the properties of the observed SHB sample, since the
range of models considered in the literature inevitably predicts a
large fraction of short-lived merging systems, which will preferably
reside in late-type hosts and occur at higher redshifts, in contrast
with the observed SHB sample which is dominated by events located in
early-type hosts and at lower redshifts.

In this work we apply a refined theoretical analysis to the
\citet{gal-yam05} sample in order to constrain  the local rate and
the lifetime of SHB progenitors. Our goal is to confront our
findings with compact binary models for SHBs and discuss the
implication for gravitational wave detection from SHBs by the LIGO
and Virgo observatories.

The structure of the paper is as follows: We develop the theory that
we need for the analysis of the data in \S2. The observed sample,
its analysis and the derived constraints are presented in \S3. We
compare our results to the predictions of compact binary mergers in
\S4 and discuss the prospects for detecting GWs from SHBs under the
assumption that they originate from a DNS or BH-NS mergers in \S5.
We summarize our conclusions in \S6.

\section{Theory}\label{SEC theory}

We use the combination of the two-dimensional {\it observed}
redshift and luminosity distribution together with the observed SHB
flux distribution (logN-logS), to constrain the {\it intrinsic} SHB
rate at redshift $z$,  $R_{SHB}(z)$ and the intrinsic luminosity
function $\phi(L)$. This method is an extension of previous works
\citep{Piran92,Ando04,Guetta05} that used only the one-dimensional
observed redshift distribution and the peak flux distribution.
Although we focus here on SHBs, the method is applicable to
long-soft GRBs as well.

The two-dimensional observed redshift and luminosity distribution is
derived from the intrinsic distributions via:
\begin{equation}\label{EQ NLz}
\frac{d\dot{N}_{obs}}{dLdz} = \phi(L)
\frac{R_{SHB}}{1+z}\frac{dV}{dz} S(P),
\end{equation}
where $\dot{N}_{obs}$ is the {\it observed} SHB rate and $\phi(L)$
is the {\it intrinsic} peak luminosity function (which we assume to
be independent of z). $0<S(P)<1$ is the probability for detection
(including redshift determination) of a burst with a peak photon
flux $P$ which in turn depends on $L$ and $z$ as well as on the
spectrum of the bursts. $R_{SHB}(z)$ is the intrinsic SHB rate per
unit comoving volume and comoving time. Since SHB progenitors are
most likely of a stellar origin we expect:
\begin{equation}\label{EQ R_SHB}
    R_{SHB}(z) \propto \int_z^\infty SFR(z')f\left(t(z)-t(z')\right)
    \frac{dt}{dz'} dz',
\end{equation}
where $SFR(z)$ is the star formation rate at redshift $z$ (per unit
comoving volume and comoving time), $t(z)$ is the age of the
universe at redshift $z$, and $f(\tau)$ is the fraction of SHB
progenitors that are born with a lifetime $\tau$.

$S(P)$ can describe a single detector or a combination of several
detectors, each weighted by its field of view and operational time.
In principle if $S(P)$ is well known and if the observed sample is
large enough then the  intrinsic distributions can be extracted from
Eq. \ref{EQ NLz}. In reality  we have to work with a limited sample
as well as poorly understood $S(P)$. In the case of GRBs (long and
short) there is a large sample of bursts, observed by
BATSE\footnote{http://www.batse.msfc.nasa.gov/batse/}, for which
only the peak flux distribution is available while the redshift (and
thus luminosity) is unknown. The BATSE sample can constrain the
intrinsic distributions by considering the observed flux
distribution which is an integration of Eq. \ref{EQ NLz}:
\begin{equation}\label{EQ logNlogS}
    \frac {d\dot{N}_{obs}}{dP}=\frac{d}{dP}\int_0^\infty dz \int_{L,min(z,P)}^\infty dL
    \frac{d\dot{N}}{dLdz},
\end{equation}
where
\begin{equation}\label{EQ Lmin}
    L_{min}(z,P)=4\pi d_L^2 k(z) P.
\end{equation}
$d_L(z)$ is the luminosity distance and $k(z)$ is  It depends on the
spectrum of the bursts and includes the k-correction as well as the
conversion from energy flux to photon flux.  $k(z)$ is assumed to be
a function of the redshift only. Note that the LogN-LogS
distribution is $\dot{N}(>P)$ in our notation.

If $\phi(L)$ is a single power-law, $\phi(L) = \phi_0 L^{-\beta}$,
with no upper or lower cutoff (within a luminosity range that we
discuss below) then the integral over $z$ in Eq. \ref{EQ logNlogS}
does not depend on $P$ and thus the observed peak flux distribution
does not depend on $R_{SHB}$ and simply satisfies:
\begin{equation}\label{EQ SPL_LogNLogS}
   \frac {d\dot{N}_{obs}(P)}{dP} \propto P^{-\beta} S(P).
\end{equation}
Similarly, the integral over $L$ in eq. \ref{EQ NLz} results in:
\begin{equation}\label{EQ Nz}
  \frac {d\dot{N}_{obs}}{dz} = (4\pi d_L^2 k(z))^{1-\beta} \phi_0 \frac{R_{SHB}}{1+z}\frac{dV}{dz}
\int P^{-\beta} S(P) dP,
\end{equation}
thereby eliminating the dependence on the detector thresholds.
Naturally for $\beta < 2$ an upper cutoff must exist while for
$\beta>1$ a lower limit is necessary. However, if the lower cutoff
is low enough so that it affects only a negligible volume, and if
the upper cutoff is high enough so it affects only the detection at
high redshift, then Eqs. (\ref{EQ SPL_LogNLogS}) and (\ref{EQ Nz})
are applicable (these cut-offs also prevent the integral over $P$ in
Eq. \ref{EQ Nz} from diverging). Therefore, if the observed peak
flux distribution can be fitted by Eq. \ref{EQ SPL_LogNLogS} then
the luminosity function can be a single power-law. In this case we
can readily use data sets for which $S(P)$ is not well known. In
such cases Eq. \ref{EQ Nz} enables a comparison of the
one-dimensional observed redshift distribution with model
predictions. Unfortunately the {\it observed} luminosity
distribution depends on $S(P)$ even when the luminosity function is
a single power-law. If $S(P)$ is well known, a better constraint on
the intrinsic distributions can be obtained by a comparison with the
two-dimensional luminosity-redshift distribution (Eq. \ref{EQ NLz}).

The above formalism is applicable to any astrophysical transient as
long as its detectability depends only on its peak flux. Eqs.
\ref{EQ NLz} and \ref{EQ logNlogS} are the most general, assuming
only that the luminosity function does not evolve with the redshift.
Eqs. \ref{EQ SPL_LogNLogS} and \ref{EQ Nz} are applicable only when
the luminosity function is a single power-law and $k(z)$ depends
only on the redshift.

\section{Constraints on progenitor lifetime and the rate of
SHBs}\label{SEC lifetime_constraints}

In this section we use the methods presented in \S\ref{SEC theory}
to analyze the observed sample of SHBs. We base our study on the SHB
sample compiled by \citet{gal-yam05}. This sample includes 4
relatively recent SHBs localized by {\it Swift} and {\it HETE-2},
and 4 ``historical'' bursts localized by the IPN over the years, and
studied in retrospect by these authors. We summarize the known
redshift and host galaxy properties of each burst, as compiled by
\citet{gal-yam05}, in Table 1.

\begin{deluxetable}{lllll}

%\tabletypesize{\scriptsize}
\tablecaption{Host galaxies and redshifts of SHBs} \tablewidth{8cm}
\tablehead{
\colhead{SHB} & \colhead{Redshift} & \colhead{Host Galaxy} & Association  & Ref.\\
              & $z$                & Type                  & significance &
} \startdata

790613 & 0.09 & E/S0 & $\sim 3\sigma$ & 1 \\
000607 & 0.14 & Sb   & $\sim 2\sigma$ & 1 \\
050509b & 0.22 & E/S0 & $3-4\sigma$ & 2\\
050709 & 0.16 & Sb/c &  Secure & 3\\
050724 & 0.26 & E/S0 & Secure & 4 \\
050813 & 0.72$^{\dagger}$ & E/S0 & - & 5 \\
\tableline
001204 & $>0.25[0.06]$ & - & $1[2]\sigma$ & 1 \\
000607 & $>0.25[0.06]$ & - & $1[2]\sigma$ & 1 \\
\enddata
\tablecomments{1. \cite{gal-yam05} 2.
\cite{bloom05,kulkarni05,Castro05a,Gehrels05} 3. \cite{Fox05} 4.
\cite{Berger05,Prochaska05} 5.
\cite{Gladders05,berger05b,Prochaska05}
\newline $^\dagger$ while the paper was in the referring process,
\cite{Berger06}suggested that this burst might be actually
associated with a cluster at a redshift $\approx 1.8$}
\end{deluxetable}

When using this sample for statistical studies of the observed
redshifts, as we do here, one has to account for selection biases.
Specifically, a selection bias that disfavors the identification of
the host galaxy of an SHB at high redshift may skew our results.
Fortunately, the sample we use is almost complete, making correction
for selection biases simple. At the time that the analysis was first
made {\it Swift} had detected and made prompt follow- up
observations of only four SHBs. Three are included in our sample
while the last (SHB 050906) might be associated with a galaxy at a
distance of 140 Mpc \citep{Levan05}. If correct, this low redshift
strengthens our results, as we shall see below; however, we do not
include this burst due to the association being unconfirmed at this
time  \footnote{During the refereeing process {\it Swift} localized
adittinal handful of SHBs. For most of these bursts there is
currently no redshift or host information. For SHB 051221 the
redshift and host are known \citep{Soderberg06}. We stress that this
burst, or future  bursts, cannot be added to a sample that is used
in a statistical comparison to the observations without proper
consideration of the selection effects that prevented the redshift
determination of the rest of the localized SHBs} .

The 4 IPN SHBs studied by \cite{gal-yam05} comprise a complete
sub-sample of IPN SHBs localized to within $10 ~\rm arcmin^2$,
defined by these authors based on a-priori technical issues (e.g.,
galactic latitude) which should not bias the physical properties of
these bursts. Of these 4 bursts, \cite{gal-yam05} determine probable
redshifts and host galaxy types for two events. For the other two
events they set a lower redshift limit. This limit is based on a
null detection of luminosity overdensities in the fields of these
SHBs, and thus no host galaxy information is available. Since
\citealt{gal-yam05} used a luminosity-based test for associating
hosts with SHBs, their search may be biased against bursts in
low-lumiosity galaxies, and against bursts at high redshift.  To
avoid the bias mentioned above our statistical analysis is based on
the density distributions derived from the lower limits on the
redshift of these two bursts (see \S\ref{SEC lifetime}).

We use the observed peak flux sample in the current BATSE
catalog\footnote{http://www.batse.msfc.nasa.gov/batse/grb/catalog/current/}.
Therefore it is convenient to define the flux and intrinsic
luminosity quantities discussed in \S\ref{SEC theory} in an energy
range that corresponds to the BASTE window for detection,
$50-300$~keV. In this window the spectrum can be well approximated
as a single power law $F_\nu \propto \nu^{-\alpha}$ where values
range between $-1 < \alpha < 0.5$ \citep{Ghirlanda04} with a typical
value $\alpha \approx -0.5$. Using this spectrum we obtain $k(z)
\approx 2 \times 10^{-7} (1+z)^{-(1-\alpha)} ~\rm erg$.

\subsection{The progenitor lifetime of SHBs}\label{SEC lifetime}

\subsubsection{Single power-law luminosity functions}\label{SEC
single_PL}

Following the discussion in $\S~$\ref{SEC theory} we first try to
fit the observed peak flux distribution $\frac{d\dot{N}(P)}{dP}$
using a single power-law with an index $\beta$. From the current
BATSE catalog we extract all the short ($T_{90}<2$s) bursts with
peak flux in the $64$ ms timing window of $P_{64}>1.5 ~\rm
ph/cm^2/sec$ (at this range $S(P) \approx 1$ for BATSE), resulting
in a list of 340 bursts. We calculate the maximum likelihood and
find that $\beta = 2 \pm 0.1$ provides the best fit. The
$\chi^2/d.o.f$ of this model is 1.17, confirming it is indeed a good
fit. Using Eq. \ref{EQ SPL_LogNLogS}, we therefore initially
consider a single power-law luminosity function $\phi(L) \propto
L^{-2}$.

The $8$ SHBs with known (or constrained) redshifts included in our
sample were detected by several different instruments (see Table 1).
The detection threshold for each experiment is not well known and
construction of $S(P)$ for this sample is therefore currently
impossible. Hence, the full sample can be used only for comparison
with models of the redshift distribution based on a single power-law
luminosity function. Should a larger sample of bursts for which the
detector sensitivity is well understood become available in the
future, the more constraining comparison with model predictions
using the two-dimensional $L$-$z$ distribution (Eq. \ref{EQ NLz})
could be carried out.

Following \citet{Schmidt01} and \cite{Guetta05} we base our redshift
distribution model calculations on the star
formation history (SFH) parametrization SF2 of \citet{Porciani01}:
\begin{equation}\label{EQ SFH2}
   SFR_2(z)\propto \frac{{\rm exp}(3.4 z)}{{\rm exp}(3.4 z) + 22}\frac{[\Omega_m(1 + z)^3 +\Omega_\Lambda]^{1/2}}{(1 + z)^{3/2}}
\end{equation}
and we adopt the standard cosmology ( $\Omega_m = 0.3$,
$\Omega_\Lambda = 0.7$ and $H_0=70 ~\rm km s^{-1}~ Mpc^{-1}$). The
probability distributions  we consider for the lifetime  are
lognormal, $f(\tau)d\tau= (\tau\sigma\sqrt{2\pi})^{-1}
exp[-(ln(\tau)-ln(\tau_*))^2/2\sigma^2] d\tau$ with various values
of $\tau_*$ and narrow ($\sigma=0.3$) or wide ($\sigma=1$)
dispersions, as well as power-law distributions $f(\tau) \propto
\tau^{\eta}$ with a lower cutoff at $20$ Myr and an upper cutoff
that is larger than the Hubble time (our results do not depend on
the exact values of the cutoffs).  For each of these models we carry
out a maximum likelihood analysis \citep{Press92} in order to
constrain the typical delay ($\tau_*$) or power-law index ($\eta$).
We do so by calculating the likelihood of the observations (table 1)
for every set of model predictions (calculated using Eqs. \ref{EQ
R_SHB}, \ref{EQ Nz} \& \ref{EQ SFH2}). The two bursts with redshift
lower limits are accounted for in the following way. For each burst
we calculate the functional form of the lower limit, l(z) which is
defined as 1 minus the significance level in which the hypothesis
that the burst is at redshift z is rejected. So, l(z) goes to 1 for
high redshifts (the possibility that these bursts are at high-z is
not constrained) and l(z) decreases with z, for example, assuming
the value 0.05 at z=0.06 (table 1). The method used to calculate
l(z) is described in detail in \cite{Nakar05} and is applied to
these bursts in \cite{gal-yam05}. The likelihood of a burst with a
lower limit is the probability that an observed burst will be
consistent with the observed lower limit for a given model of
redshift distribution: $\int (d\dot{N}_{obs}/dz) l(z) dz $.

\begin{figure}
\includegraphics[width=8cm]{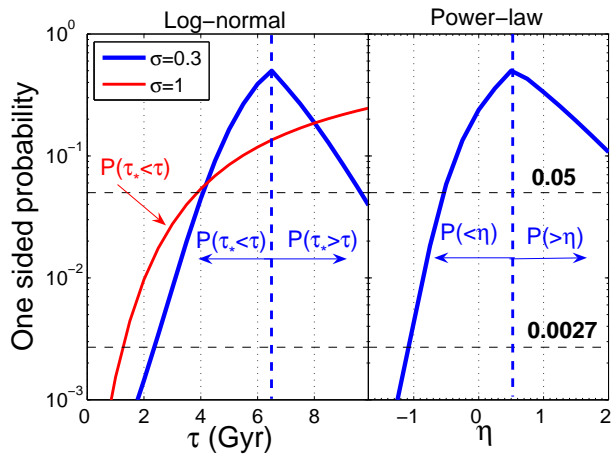}
\caption{{\bf Left}: The one-sided probability curves, which result
from the maximum-likelihood analysis, for a narrow ({\it thick
line}) and wide ({\it thin line}) lognormal lifetime distributions.
For lifetime values, $\tau$, that are smaller [larger] than the most
probable value, the one-sided probability is the probability that
$\tau_*<\tau$ [$\tau_*>\tau$]. For the most probable value this
probability is 0.5. Considering a narrow lognormal distribution the
most likely delay is $\tau_* = 6.5$ Gyr and its $5\%-95\%$
confidence interval is $4<\tau_*<9.5$ Gyr, while the probability
that $\tau_*>1$ Gyr is $99.98\%$. Assuming a wide lognormal
distribution the most likely delay is $\tau_* = 20$ Gyr while
$\tau_* >4 [1]$ Gyr at $95[99.9]\%$. {\bf Right}: The one-sided
probability curves of a single power-law lifetime distribution. The
most probable power-law index is $\eta=0.6$, the $5\%-95\%$
confidence interval is $-0.5<\eta<2.6$ and the probability that
$\eta>-1$ is $99.5\%$. See \S\ref{SEC single_PL} for
details}\label{FIG probability_curves}
\end{figure}

The results of the likelihood analysis that we carry for the
different lifetime distributions are presented as one-sided
probability curves in Figure \ref{FIG probability_curves}. For a
model likelihood $\cal{L}$ the one-sided probability is defined as
$(1-P_{\chi^2,1}(2ln[{\cal L}_{max}/{\cal L}]))/2$ where ${\cal
L}_{max}$ is the maximal likelihood and $P_{\chi^2,1}$ is the
cumulative $\chi^2$ distribution with one degree of freedom (e.g.,
\citealt{Press92}). Figure \ref{FIG probability_curves} shows
clearly that long life-times are favored. For the narrow lognormal
distribution we find that the most likely delay is $\tau_* = 6.5$
Gyr and its $5\%-95\%$ confidence interval is $4<\tau_*<9.5$ Gyr,
while the probability that $\tau_*>1$ Gyr is $99.98\%$. Assuming a
wide lognormal distribution we find that the most likely delay is
larger than the Hubble time ($\tau_* = 20$ Gyr), that $\tau
>4$ Gyr at $95\%$ confidence and that the probability that $\tau_*>1$ Gyr
is $99.9\%$. Considering a power-law lifetime distribution we find
that the most probable power-law index is $\eta=0.6$, the $5\%-95\%$
confidence interval is $-0.5<\eta<2.6$ and the probability that
$\eta>-1$  is $99.5\%$ \footnote{ In these calculations we took the
redshift of SHB 050813 as 0.72
\citep{Gladders05,berger05b,Prochaska05}. while the paper was in the
refereeing process, \cite{Berger06} suggested that this burst might
actually be associated with a cluster at a redshift higher than
0.72. Repeating the analysis while taking a lower limit of $z>0.72$
for SHB 050813 does not change the results significantly. For
example, for narrow lognormal distribution, the most likely delay is
$\tau_* = 5.8$ Gyr, the $5\%-95\%$ confidence interval is
$3.5<\tau_*<9$ Gyr while the probability that $\tau_*>1$ Gyr is
$99.8\%$.}.

\begin{figure}
\includegraphics[width=8cm]{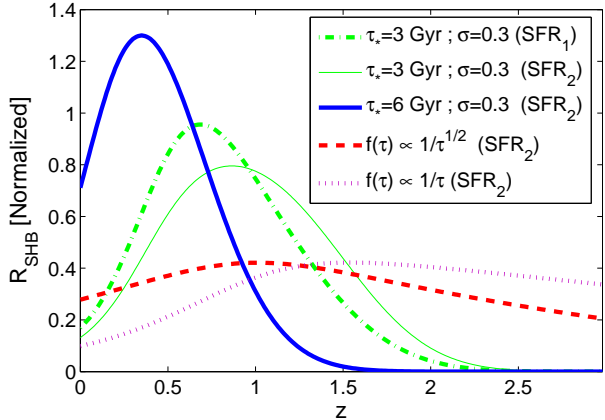}
 \caption{The normalized {\it intrinsic} SHB rate ($R_{SHB}$) as a function of the redshift for several
 lifetime distributions and SFH parameterizations (see legend).
}\label{FIG R_SHB}
\end{figure}

\begin{figure}
\includegraphics[width=8cm]{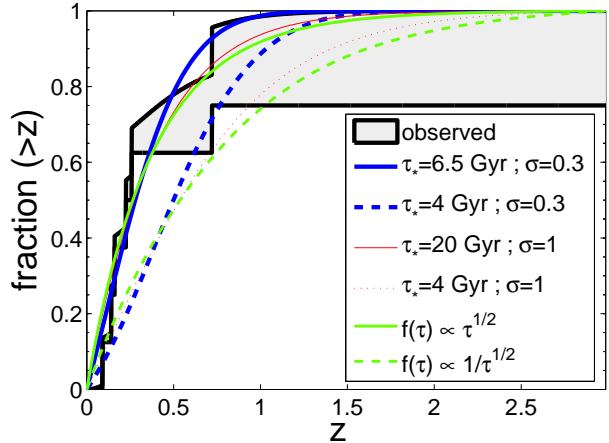}
 \caption{ The cumulative {\it observed} redshift distribution as predicted
 by various lifetime distributions when the luminosity function is
$\phi(L) \propto L^{-2}$ and the star formation history is given by
Eq. \ref{EQ SFH2}. For each functional form of the lifetime
distribution we present the most likely distribution and the
distribution that is rejected at $95\%$ confidence level. The
cumulative redshift distribution of the observed data (shaded area)
is bracketed between the lower solid line, which is the cumulative
redshift distribution of the six bursts with known redshifts, and
the upper solid line, which includes also the contribution of the
two bursts with upper limits, as given by l(z) (see text). The
figure demonstrates  that the most likely models provide a good fit
to the data while the models that are rejected under-predict the
fraction of bursts at low redshift $z\lesssim 0.3$}\label{FIG Zdist}
\end{figure}

The normalized ({\it intrinsic})redshift distributions,
$R_{SHB}(z)$,  resulting from various lifetime distributions are
presented in Figure \ref{FIG R_SHB}. Comparison between the
predicted and observed cumulative redshift distributions for several
representative models are depicted in Figure \ref{FIG Zdist}. This
figure vividly illustrate the results from our maximum likelihood
analysis, namely that models  consistent with the data must have a
long typical delay($\gtrsim 4$ ~\rm Gyr). Models with no typical
lifetime (power-law distributions) must have a birthrate of
progenitors per unit logarithmic lifetime that increases
significantly as a function of lifetime (i.e. $\eta>-0.5$). This
figure also shows  that indeed all the models we considered provide
a good fit to the data when $\tau_*$($\eta$) takes its most likely
value. It also demonstrates that the reason that models dominated by
short-lived systems are rejected is that they under-predict the
fraction of bursts at low redshift ($z\lesssim 0.3$).

Given that the sample that we use is small, and to test for biases
in our maximum likelihood analysis, we have carried out a jackknife
analysis \citep{Efron82}. Namely, we repeat the analysis 8 times, in
each occasion  removing a different burst from the sample (i.e.
using a different sample of 7 bursts). We find a negligible bias,
for example the bias of the most likely $\tau_*$ [$\eta$] for a
narrow lognormal [power-law] distribution is $-0.05$ Gyr [$0.15$],
implying that the most likely values that are obtained by the
maximum likelihood analysis are not biased because of the small
sample size. We also find that our results are not driven by any
single burst. For example, for a narrow lognormal lifetime
distribution the most likely value of $\tau_*$ in the different
samples varies between 5.8 Gyr and 8 Gyr (in the full sample it is
$6.5$ Gyr) and in all the samples $\tau_*>3.4$ Gyr at more than
$95\%$ confidence level while the probability that $\tau_*>1$ Gyr
is, in all samples, higher than $>99.8\%$.

Given the observational uncertainty in the SFH at high redshift we
repeat our analysis using the SFH formula SF1 from
\citet{Porciani01}:
\begin{equation}\label{EQ SFH1}
   SFR_1(z)\propto \frac{{\rm exp}(3.4z)}{{\rm exp}(3.8z) + 45}\frac{[\Omega_m(1 + z)^3 +\Omega_\Lambda]^{1/2}}{(1 +
   z)^{3/2}}.
\end{equation}
In this model the star formation rate falls exponentially at
redshifts larger than 1.5. The results we obtain using this SFH
model are similar to those obtained using Eq. \ref{EQ SFH2}. For
example considering a narrow lognormal distribution with $SFR_1$ we
find that the most likely delay is $\tau_* =6$ Gyr and its
$5\%-95\%$ confidence interval is $3.5<\tau_*<9$ Gyr while the
probability that $\tau_*>1$ Gyr is $99.95\%$. The reason is that the
models differ only at high redshift, with little impact on the
distribution at low redshifts. Of the three models presented in
\citet{Porciani01} this model places the most conservative
constraints on $f(\tau)$ as it predicts the largest fraction of SHBs
at low redshift (for a given lifetime distribution). Therefore,
lifetime distributions that are ruled out when applied to this model
will be ruled out with greater significance when the other
\citet{Porciani01} SFH models are used.

We return now to the upper and lower cutoffs of the luminosity
function. As described above, for a given selection of $R_{SHB}$ and
$S(P)$ there are critical values of the lower (upper) luminosity
function cutoffs, below (above) which the observations are
insensitive to the value of the cutoff. Among all the models of
$R_{SHB}$ that we have examined, and for threshold values that vary
between\footnote{This range includes the estimated thresholds of all
the instruments that contributed to the detection of our SHB
sample.} $1-50~ \rm ph/cm^2/s$, the fit to the LogN-logS
distribution, as well as the fit to the observed redshift
distribution, are insensitive to a lower cutoff $L_{min}<10^{49}$
erg/s and to an upper cutoff $L_{max}>10^{52}-10^{53}$ erg/s.
Namely, a luminosity function with $\phi(L)\propto L^{-2}$ for
$10^{49} <L<10^{53}~\rm {erg/s}$ and $\phi(L)=0$ otherwise is
consistent with all available data. The luminosity of every SHB in
our sample falls within this range as well. Although the sample that
we consider is consistent with $L_{min}=10^{49}$ erg/s, recent
results suggest a lower $L_{min}$. First, if SHB 050906 is
associated with the nearby galaxy IC 328 at $140$ Mpc, as suggested
by \citet{Levan05}, then its luminosity is about $10^{48}$erg/s.
Second, \citet{tanvir05} found evidence that at least $5\%$ of the
SHBs detected by BATSE are at $z<0.025$ ($\approx 100$ Mpc).
Interestingly, a comparable fraction ($\approx 5\%$) is predicted by
our best fit model ($\phi(L) \propto L^{-2}$ and a typical lifetime
of $6$ Gyr) if the lower cutoff is as low as $L_{min}<10^{46}$
erg/s. A slightly smaller, but consistent, fraction ($\approx 3\%$)
is predicted if $L_{min}=10^{47}$ erg/s while the fraction drops to
less than $0.5\%$ for $L_{min}=10^{48}$ erg/s.

If $L_{min} \lesssim 10^{47}$ erg/s then the SHB luminosity function
overlaps with the energies observed in giant flares from soft
gamma-ray repeaters (SGRs) \citep{Hurley05,Palmer05}. This overlap
in luminosities raises (again) the question whether SGRs can produce
flares with isotropic equivalent observed luminosities much greater
than $\sim 10^{47}$ erg/s, as we previously suggested in
\citet{Nakar05}. In this case giant flares from SGRs can be the
source of all observed SHBs \citep{Dar05a}. The fact that SHBs are
observed in elliptical galaxies cannot a priory rule out this
possibility since it might be that neutron stars (and thus SGRs) can
still be produced long after star formation ended in such galaxies
\citep{Dar05b}. For example, it has been proposed that neutron stars
can be the end product of white dwarf binary mergers \citep{Saio85}
or of an accretion-induced collapse of a white dwarf
\citep{Nomoto91}. However, observations indicate that SGRs and
anomalous x-ray pulsars (AXPs) -- two known populations of magnetar
candidates in our Galaxy -- are associated with star forming regions
\citep{Gaensler01}. Thus our result that SHBs are dominated by
progenitors with lifetime of several Gyr disfavor this possibility.

\subsubsection{Broken power-law luminosity functions}

Although a broken power-law luminosity function is not necessary in
order to explain the data, this, or even more complicated functional
forms, cannot be excluded. Given the prominence of broken power-law
luminosity functions in previous studies we also consider a
luminosity function of the form:
\begin{equation}\label{EQ BPL_LF}
    \phi(L) \propto \left\{\begin{array}{c}
                             L^{-\alpha_1} ~~~L<L_* \\
                             L^{-\alpha_2} ~~~L>L_* \\
                           \end{array} \right.  ,
\end{equation}
where $\alpha_1<1.5$ and $\alpha_2>2.5$. We choose this range of
indices so the luminosity function will significantly deviate from
the single power-law for which we already obtained a constraint on
the lifetime. Another reason for exploring a broken power-law
distribution is that it demonstrates the analysis that should be
done once the sample of {\it Swift} short bursts with known redshift
will be significantly larger and the {\it Swift} threshold will be
better understood. Such luminosity functions couple the intrinsic
functions to the detector response when Eq. \ref{EQ NLz} is
integrated over $L$ or $z$. Therefore, in this case, $S(P)$ needs to
be understood in order to compare the observations with the model.
Since we cannot construct $S(P)$ for the whole sample we use in this
section only the three bursts detected by {\it Swift}. However, as
we now account for $S(P)$ (and its uncertainties), we can use the
two-dimensional distribution (Eq. \ref{EQ NLz}). Based on the peak
flux of the three {\it Swift} SHBs with known redshifts we estimate
the {\it Swift}/BAT threshold for localizations to be comparable to
that of BATSE. We therefore approximate Swift's threshold as a step
function at $1 \rm{ ~ph/cm^2/s}$ with a large uncertainty due to the
small number of {\it Swift} SHBs. We use Eq. \ref{EQ SFH2} for the
star formation history.

In order to test the effect of broken power-law luminosity functions
on the derived limits on the lifetime we carry a maximal likelihood
analysis for the narrow lognormal and the single power-law lifetime
distributions. For each model we first fix $\alpha_1=1.5$ and
$\alpha_2=2.5$ and fit $L_*$ to the BATSE logN-logS distribution. A
narrow lognormal lifetime distribution results with a most likely
delay of $\tau_* =6.5$ Gyr and $\tau_*>4 [1]$ Gyr at $95\%
[99.98\%]$. A power-law lifetime distribution results with a most
likely power-law index of $\eta = 1$ and with $\eta>0 [-1]$ at $95\%
[99.9\%]$. Repeating the analysis with other values of $\alpha_1$
and $\alpha_2$ (fitting $L_*$ to the BATSE logN-logS distribution)
we obtain similar results. Given the uncertainty in the {\it Swift}
threshold, we have also varied the assumed threshold value by a
factor of 5 below our best estimate (retaining its step function
form), again finding similar results. Note that although here we can
use a significantly smaller sample (3 bursts) than the one we used
in \S\ref{SEC single_PL} (8 bursts), we obtain similar significance.
The reason is, at least partially, the usage of the two-dimensional
distribution which contains more information than the redshift
distribution alone. This sample is however much more vulnerable to
unaccounted errors\footnote{Repeating the analysis when the redshift
of SHB 0508013 is taken as a lower limit $>0.72$ (instead of
$=0.72$) does not alter the main results, that a long lifetime is
required. The most likely values of $\tau_*$ and $\eta$ are similar
while $\tau_*<1$ Gyr is rejected at $99.7\%$ and $\eta <-1$ is
rejected at $99.4\%$}.

\begin{figure}
\includegraphics[width=8cm]{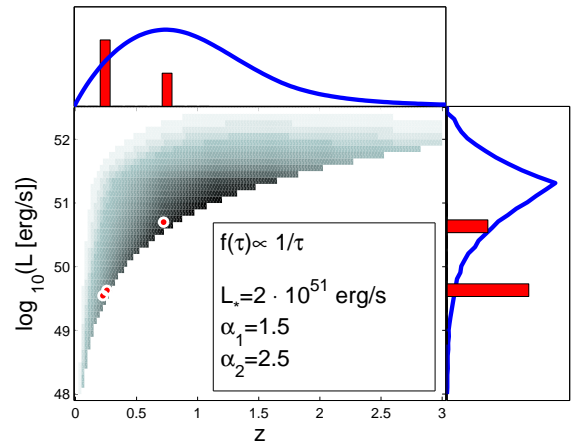}
\caption{A color plot of the two-dimensional distribution
$\frac{d\dot{N}_{obs}}{dlog(L)dz}$ and its one dimensional
projections (obtained by integrating over $z$ or $log(L)$) as
predicted by a model with a lifetime distribution $f(\tau) \propto
1/\tau$ and an SFH of the form given by Eq. \ref{EQ SFH2}. The
luminosity function is a broken power-law with $\alpha_1=1.5$ and
$\alpha_2=2.5$ and $L_*=2 \cdot 10^{51} ~\rm erg/s$, chosen to fit
the BATSE LogN-LogS distribution. The threshold of {\it Swift} is
taken as a step function at  $1 \rm{ ~ph/cm^2/s}$. The Three Swift
SHBs are marked with dots (on the two-dimensional plot) and bars (in
the two projections). This model is rejected at $99.9\%$ confidence
level.}\label{FIG NZdist1}
\end{figure}

\begin{figure}
\includegraphics[width=8cm]{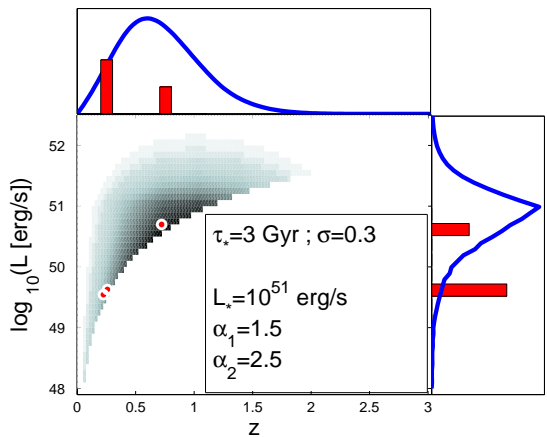}
 \caption{Same as Fig. \ref{FIG NZdist1} for a model with a lognormal
 lifetime distribution ($\tau_*=3 ~\rm Gyr$ and $\sigma=0.3$)
 and $L_*=10^{51}$ erg/s (the rest of the parameters are the same as in
 Fig. \ref{FIG NZdist1}). This model is rejected at $99\%$ confidence level.
  }\label{FIG NZdist2}
\end{figure}

Figures \ref{FIG NZdist1} and \ref{FIG NZdist2} depicts the
predicted two-dimensional distributions of a power-law lifetime
distribution with $\eta=-1$ and a narrow lognormal lifetime
distribution with $\tau_*=3$ Gyr, respectively. These figures
include also the observed distribution as well as the two
one-dimensional projections of the models and the observations. The
figures show clearly that even though there are only three observed
SHBs in this sample the two models do not describe the data well.
Indeed both models are rejected at a confidence level that is larger
than $99\%$.


\begin{figure}
\includegraphics[width=8cm]{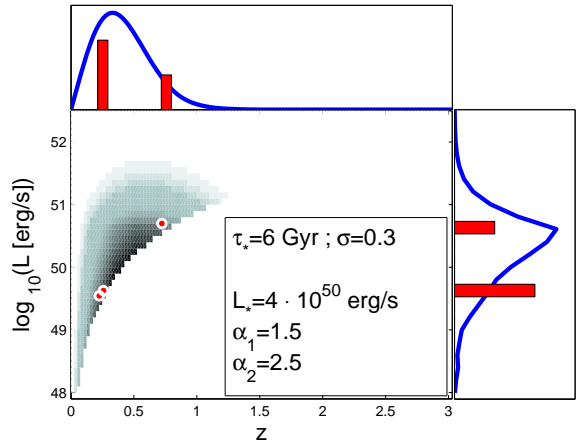}
 \caption{Same as Fig. \ref{FIG NZdist1}, for a model with a lognormal
 lifetime distribution ($\tau_*=6 ~\rm Gyr$ and $\sigma=0.3$)
 and $L_*=4 \cdot 10^{50}$ erg/s (the rest of the parameters are the
 same as in Fig. \ref{FIG NZdist1}). This model provides a good description
 of the data.  }\label{FIG NZdist3}
\end{figure}

This result can be understood. Given the intrinsic rate evolution
$R_{SHB}(z)$ predicted by lifetime distributions with short delay
($<3$ Gyr) or small power-law index ($\eta<-0.5$), the luminosity
function (Eq. \ref{EQ BPL_LF}) implies that the typical observed
burst luminosity is $L_*$, and the typical observed redshift is the
redshift at which $L_*$ is detected at the threshold
level\footnote{This statement is correct as long as $L_*$ is
detected up to a redshift $\lesssim 1$, which is the case for the
values of $L_*$ that we find and the {\it Swift} threshold.}.
Regardless of the exact value of $\alpha_1$ and $\alpha_2$ we find a
typical luminosity $L_* \gtrsim 10^{51}$. Similar values of $L_*$
were obtained by \citet{Ando04} for various luminosity functions.
The typical redshift is $z\approx 0.7$, and together with $L_*$
these two values are too high for the {\it Swift} sample. Longer
lifetimes result in $R_{SHB}(z)$ that increases at low redshifts,
leading to a lower value of $L_*$ as well as to lower typical
redshifts. For example, a delay of $6$ Gyr results in $L_*= 4 \times
10^{50}$ erg/s and predicts a distribution that is fully consistent
with the observations (see fig. \ref{FIG NZdist3}).

Finally we discuss the robustness of our limit on the progenitor lifetime in
the case of a luminosity function of an arbitrary form. The reason
that short lifetime models fail is the deficit of low
luminosity SHBs at low redshifts. In order to
increase this fraction the luminosity function must increase more steeply
than $L^{-2}$ at low luminosities ($\lesssim 10^{50}$ erg/s). In
this case however the luminosity function must become flatter than
$L^{-2}$ at higher luminosities in order to fit the BATSE LogN-logS
(i.e the luminosity function must have an 'ankle'). Although we
cannot exclude a luminosity function with an 'ankle', and thus the
possibility of a shorter typical lifetime, such a luminosity function
is not expected unless SHBs are composed of two separate
populations.

An alternative method to estimate the progenitor lifetime is based
on the spectral types of the host galaxies \citep{gal-yam05}. In the
extended sample four out of the six putative host galaxies are of
early type \footnote{Note that the host of SHB 050813 is of
early-type regardless of whether it is associated with the cluster
at z=0.72 \citep{Gladders05,berger05b,Prochaska05} or with a cluster
at $z \approx 1.8$ \citep{Berger06}. Moreover, if the burst is truly
associated with a cluster at $z \approx 1.8$ than the lifetime of
its progenitor is $\approx 3$ Gyr. It cannot be older since this is
the age of the universe at this redshift and it is not much younger
since it is associated with one of the most developed, and therefore
one of the oldest, structures at this redshift.}. If only {\it
Swift} and {\it HETE2} bursts are considered then three out of four
are of early-type. The two late-type galaxies are moderately
star-forming \footnote{Note that as we discuss in \S\ref{SEC
lifetime_constraints} the host of SHB 051221 cannot be included in
this sample without accounting for selection effects that prevent
the detection of the hosts of other SHBs that were localized after
we defined our sample. This is especially important when the host
type is considered since we expect afterglows to be bright in a gas
rich environments, making the identification of a late-type hosts
more probable.}. In one a significant fraction of the stars is $\sim
1$ Gyr old (SHB 050709; \citealt{Covino06}). In the other most of
the stars are much older than $1$ Gyr (SHB 000607;
\citealt{gal-yam05}). Using these results \cite{gal-yam05} find that
SHBs are more likely to reside in early type galaxies than type Ia
supernovae (at $2\sigma$ c.l.), implying that the {\it typical}
lifetime of SHB progenitors is most likely longer than the one of
type Ia SNe ($>1$ Gyr). This method does not depend on the
luminosity function or the instrumental thresholds and thus it
constitutes an independent corroboration of our results.

\subsection{The Local Rate of SHBs}\label{SEC rate}

The most robust lower limit on the local SHB rate is obtained
directly from the BATSE observed rate (a full sky rate of  $\approx
170 ~\rm yr^{-1}$). Considering a single power-law luminosity
function the observed redshift distribution of the extended sample
(8 bursts) should be similar to the BATSE redshift distribution of
BATSE bursts. Therefore, the fact that 5 out of the 8 bursts are
within a distance of 1 Gpc suggests that the rate of SHBs observed
by BATSE is $\approx 25 ~\rm Gpc^{-3}~ yr^{-1}$ and puts a lower
limit of $\approx 12 [6] ~\rm Gpc^{-3}~ yr^{-1}$ at $2[3]\sigma$
confidence level. Even with no assumptions about the luminosity
function, the sample of SHBs detected by {\it Swift} (3 bursts),
which has a comparable threshold to that of BATSE, gives a similar
result. Two of these bursts are within 1 Gpc suggesting the same
rate of $\approx 25 ~\rm Gpc^{-3}~ yr^{-1}$ and setting a lower
limit of $\approx 6 [1] ~\rm Gpc^{-3}~ yr^{-1}$ at $2[3]\sigma$
confidence level. The rate of SHBs observed by BATSE is a strict
lower limit since it does not include undetected SHBs. Bursts can
avoid detection either by pointing away from us, if the prompt
gamma-ray emission is beamed, or by being too dim. This robust lower
limit is higher than previous estimates
\citep{Schmidt01,Ando04,Guetta05} by a factor of $10-100$. This
observed local rate is also higher by the same factor than the
observed local rate of long-soft GRBs \citep{Schmidt99,Guetta05b}.

The progenitors of SHBs are almost certainly a product of at least
one core-collapse supernova (e.g., a neutron star) and the SHB
itself is most likely a catastrophic event, therefore the rate of
SHBs is limited by the rate of core-collapse SNe. Since we find that
the typical lifetime of SHB progenitors is several Gyr the local
rate of SHBs is limited by the rate of  core collapse SNe at a
redshift $\approx 0.7$ \citep{Dahlen04}. Together with the observed
lower limit we find that the local rate of SHBs, ${\cal
R}_{SHB}\equiv R_{SHB}(z=0)$, is in the range:
\begin{equation}\label{EQ SNe rate}
   10 \lesssim  {\cal R}_{SHB} \lesssim 5 \times 10^5 \rm ~Gpc^{-3} ~yr^{-1}.
\end{equation}

Constraining the local rate within this range requires an estimate
of the beaming factor (i.e., the fraction of the $4\pi$ solid angle
into which the prompt gamma-rays are emitted), $f_b^{-1}$, and of
the luminosity function. The value of the beaming factor is
currently unknown, however, the afterglows of two SHBs (050709 and
050724) have shown a steepening that can be interpreted as a hint of
a jet \citep{Fox05}. This interpretation, although somewhat
speculative at this point, indicates a beaming factor of $\sim 50$.
The correction for undetected dim bursts depends on the luminosity
function and most strongly on its lower cutoff, $L_{min}$. Taking
the power-law luminosity function that we have found above ($\phi(L)
\propto L^{-2}$) the local rate is weakly sensitive to the exact
evolution of $R_{SHB}$ with redshift (as long as the fraction of
nearby bursts is compatible with the current sample \footnote{ The
rate varies by $\approx 30\%$ when calculated using the most likely
values of the three models that we consider. Adopting values values
of the models that are rejected at $2[3]\sigma$ can reduce the rate
by a factor of $3[6]$.}):
\begin{equation}\label{EQ R_SHB_Z0}
    {\cal R}_{SHB} \approx 40 f_b^{-1} \left(\frac{L_{min}}{10^{49}
    \rm{erg/s}}\right)^{-1}
    \rm ~Gpc^{-3} ~yr^{-1}.
\end{equation}
Taking the beaming factor suggested by \citet{Fox05} (30-50) and the
value of $L_{min}$ suggested by our results in conjunction with the
results of \citet{tanvir05} ($10^{47} erg/s$), we obtain  a rate of
$10^5 \rm ~Gpc^{-3} ~yr^{-1}$. This high rate is highly uncertain at
this point but we consider it to be the most favored by the
synthesis of current observations. This rate is only slightly
smaller than the upper limit we obtained (Eq. \ref{EQ SNe rate}),
implying that if this estimate is correct a significant fraction of
all the end products of core-collapse SNe are producing SHBs.
Alternatively, perhaps a source can produce more than a single SHB
(i.e. SHBs are not catastrophic events). The upper limit that we
find does not pose at the moment a stringent limit on the beaming or
on $L_{min}$. However, if in the future bursts with a luminosity of
$10^{47}$ erg/s will be observed, it will constrain the beaming to
be smaller than $\sim 100$.

%%%%%%%%%%%%%%%%%%%%%%%%%%%%%%%%%%%%%%%%%%%%%%%%%%

\section{SHBs and compact binary mergers}\label{SEC mergers}

\subsection{The predicted lifetime and merger rate of DNS and BH-NS binaries}

The merger rates of DNS and BH-NS systems are currently estimated in
two ways, based on  observed systems in our galaxy and using
theoretical population synthesis. Estimates based on observed DNS
binaries (e.g.,
\citealt{phinney91,nps91,cl95,vl96,acw99,kns+01,kkl+04,frv+05})
provide a lower limit on the local rate of DNS merger events because
of possible unaccounted for selection affects (e.g., binary
formation that does not recycle neither of the neutron stars). The
absence of observed BH-NS systems together with poor understanding
of selection affects involved in their discovery renders this method
unapplicable in this case. Population syntheses (e.g.
\citealt{Lipunov95,Portegies98,Bethe98,bsp99,Fryer99,bk01,Belczynski02a,Belczynski02b,Perna02})
do not suffer from observational selection effects and can address
both DNS and BH-NS binaries. However, the uncertainties involved are
substantial, with rate estimates spanning over three orders of
magnitude.

The estimates based on observed DNS systems were drastically revised
with the discovery of the relativistic binary pulsar
PSR~J0737$-$3039 \citep{bdp+03}. Based on three observed DNS
binaries\footnote{\citet{kkl+04} consider only the observed DNS
systems that will merge within a Hubble time. They further exclude
PSR~B2127+11C because of its association with a globular cluster.}
\cite{kkl+04} find a range of Galactic merger rates from $1.7\times
10^{-5}$ to $2.9\times 10^{-4}$\,\peryear\ at 95\% confidence. This
estimate is larger by a factor of 6-7 than the estimates that they
obtained excluding PSR~J0737$-$3039. Extrapolation of this rate to
the cosmological neighborhood \citep{kkl+04} yields \mbox{LIGO-I}
and \mbox{LIGO-II} detection rates of $7\times
10^{-3}-0.1$\,\peryear\ and $40-650$ \peryear\ , respectively. This
rate corresponds to a local rate density of $200-3000 ~\rm Gpc^{-3}
~yr^{-1}$.

Regardless of the exact values of the above estimated rate, the
merger rate for observed DNS binaries is dominated by systems with
short lifetimes. This is demonstrated, for example, by the fact that
the detection of PSR~J0737$-$3039 (with a merger timescale of
$\approx 100$\,Myr) significantly increased the estimated merger
rate \citep{kkl+04}. Note that this drastic revision in the rate is
partially due to the short lifetime of this double pulsar. Numerous
new binaries with long lifetimes need to be discovered if old
binaries are to dominate the merger rate in our galaxy.

The rates found by population syntheses for DNS systems are in
general consistent with the rate deduced from the observations.
Analysis of BH-NS binaries shows that their merger rate may be
comparable, or even larger by an order of magnitude \citep{Bethe98},
than the rate of DNS mergers. Population synthesis provides a rather
robust upper limit on the rate of DNS and BH-NS mergers by
considering the fraction of binaries that survive two core-collapse
SNe. Several theoretical works (e.g., \citealt{Pfahl02,Lipunov97})
show that this fraction is likely to be $\sim 0.001$ and can be as
high as $0.02$. Taking this fraction from the local rate of
core-collapse SNe \citep{Cappellaro99} implies an upper limit of
$1000 ~\rm Gpc^{-3} ~yr^{-1}$ on the merger rate of short-lived
binaries ($\tau \lesssim 1$ Gyr). An upper limit of $10^4 ~\rm
Gpc^{-3} ~yr^{-1}$ on the merger rate of long-lived binaries($\tau
\gtrsim 4$ Gyr) is obtained by considering the rate of core-collapse
SNe at redshift 0.7 \citep{Dahlen04}.

%%%%%%%%%%%%%%%%%%%%%%%%%%%%%%%%%%%%%%%%%%%%%%%%%%

\subsection{Comparison with SHBs}

DNS or BH-NS mergers are currently the most popular models for SHBs
(e.g., \citealt{Eichler89, Narayan92}). The proposed association is
based on the tight limits that are imposed on the progenitor by the
large energy release ($\sim 10^{50}$ erg) and the short timescales
($\sim 1~\rm ms$) observed in these bursts. Not many sources can
produce such an electromagnetic display, while an extensive
numerical effort (e.g.
\citealt{Ruffert99,Janka99,Rosswog03,Lee05,Oechslin05}) has shown
that the mergers of DNS and BH-NS systems can.

The range of SHB rates that we find (Eq. \ref{EQ R_SHB_Z0}) overlaps
the predicted rate of DNS mergers based on observations ($\sim
200-3000 ~\rm Gpc^{-3} ~yr^{-1}$). However, the range of progenitor
lifetimes we find for SHBs is inconsistent with that of the observed
DNS binaries. The reason is that the merger rate of the latter is
dominated by short-lived binaries ($\tau \sim 100$\,Myr) while SHB
progenitors are much older ($\tau \gtrsim 4$\,Gyr). We do not see
any obvious selection effect that may operate and can be applied to
the current DNS models and explain this discrepancy. On the
contrary, while very tight DNS binaries (with very short lifetimes)
can avoid detection in pulsar surveys because of their large orbital
accelerations, long-lived systems are detected with relative ease as
long as one of the neutron stars is an active pulsar.

Admittedly, the probability that both members of a DNS binary are
dead pulsars (i.e., undetectable), increases with its age. However,
our limits on the lifetime distribution of SHBs constrain it to be
shallower than $\tau^{-0.5}$ (at $95\%$ confidence level). In other
words, the number of newly born SHB progenitors with a lifetime
between $1-10$ Gyr is larger by a factor $\gtrsim 10$ than those
with lifetime $10-100$ Myr. Therefore, if DNS binaries are the
progenitors of SHBs, for every recently born DNS with a lifetime
$10-100$ Myr we expect to observe at least several recently born
systems with a lifetime of $1-10$ Gyr. All of these coeval systems
are presumably equally detectable via pulsar surveys. In reality
there is one observed DNS with a lifetime of $10-100$ Myr and only
one with a lifetime of $1-10$ Gyr (B1534+12, which is most likely
older than 100 Myr). We therefore conclude that with the caveat of
small number statistics, the observed population of DNS systems is
unlikely to be the dominant component among SHB progenitors.

It is possible that known DNS pulsars are not representative of the
larger population of relativistic DNS systems. This might be the
case if large numbers of relativistic DNS systems are formed without
recycled pulsars,  or alternatively if the Milky-way DNS population
do not represent the cosmic one. We thus conclude that SHB
observations disfavor a DNS merger origin to the extent that they
indicate that an unexpected hidden population of old long-lived
systems should be invoked. While we cannot exclude the existence of
such a population at the moment, it remains to be seen whether it
can be reproduced by population synthesis models. Turning the
argument around, if DNS binaries are the progenitors of SHBs then
there exists a large population of hitherto-undetected old DNS
systems. Moreover, this scenario suggests that the rate of SHBs is
${\cal R} \sim 10^3-10^4 ~\rm Gpc^{-3}~yr^{-1}$ in order to dominate
over the the merger rate of young DNS systems found by
\cite{kkl+04}, while conforming with the upper limit imposed by the
rate of core-collapse SNe at $z \approx 0.7$ and the binary survival
probability. It also excludes the existence of the
recently-hypothesized dominant population of DNS binaries with very
short lifetimes \citep{bk01,Perna02}.

The possibility that BH-NS systems are the dominant source of SHBs
cannot be constrained at this time. There are no observational data
on the properties of this population while the theoretical models
are highly uncertain and their predictions vary significantly from
one model to the next. If a link between SHBs and BH-NS mergers will
be established in the future, the observed SHB properties can be
used to constrain models of NS-BH binaries.

%%%%%%%%%%%%%%%%%%%%%%%%%%%%%%%%%%%%%%%%%%%%%%%%%%

\section{Prospects for gravitational wave detection}

Progenitors of SHBs may also be strong sources of gravitational
waves (GWs). Therefore, the local SHB rate might has direct
implications for the detection rate of GW telescopes in general and
to the upcoming S5 run of initial LIGO (LIGO-I) in particular. The
most promising progenitors for GW detection are NS-BH or DNS
mergers. As we discuss above our results disfavor, but cannot
exclude, DNS binaries as the source of SHBs, while the possibility
that NS-BH binaries are the progenitors on SHBs cannot be
constrained by the data. Moreover, energy and time scale
considerations favor compact binary mergers as the progenitors of
SHBs. Therefore it is worthwhile to discuss the predictions for the
LIGO and VIRGO GW observatories if SHBs originates during such
compact mergers.

These sources are expected to be detected by LIGO-I up to a distance
of $\sim 43[20] ~\rm Mpc$  (\citealt{Cutler02}; hereafter the
numbers without brackets are for NS-BH mergers assuming a BH mass of
$10 \rm M_\odot $ while the numbers within the brackets are for DNS
mergers\footnote{Recently \cite{Miller05} and \cite{Rosswog05}
argued that BH-NS mergers with a mass ratio $\gtrsim 10$ cannot
produce the accretion disk that is believed to be necessary for the
production of SHBs. This result suggests that SHBs can be produced
only by mergers of BH and NS with similar masses. In this case the W
detectability of BH-NS mergers is similar to that of DNS mergers.}).
VIRGO design sensitivity
\footnote{http://www.virgo.infn.it/senscurve/} is similar to that of
LIGO-I, and therefore the VIRGO detection rate is expected to be
similar once its design sensitivity is achieved. If SHBs are
triggered by compact mergers, the prospects for GW detection from
SHBs are quite promising. A reasonable, but speculative, SHB rate of
$\sim 10^4 ~\rm Gpc^{-3}~ yr^{-1}$ (see \S\ref{SEC rate}), predicts
a detection rate of $\sim 3 [0.3]$ mergers per year. This
speculative rate is higher by a factor of $\sim 10$ than previous
upper limits on merger rates derived from the local rate of
core-collapse SNe. The reason is that the long lifetime of SHB
progenitors relates them to SNe at redshift $\sim 1$, which was
higher by a factor of $\sim 10$ than it is today.

The possible temporal coincidence between an SHB and a GW signal,
which may make the difference between detection and non-detection in
the coming S5 run of LIGO-I. A coincident detection increases the
range of LIGO-I and VIRGO by a factor of $\approx 1.5$, while if
SHBs are preferentially beamed perpendicular to the binary orbital
plane the range is increased by an additional factor of $\approx
1.5$ to a distance of $\approx 100[50] ~\rm{Mpc}$
\citep{Kochanek93}. Hence LIGO-I has a non-negligible probability to
detect GWs simultaneously with a {\it Swift} detection of an SHB.
Our best fit model to the current observations ($\phi(L) \propto
L^{-2}$, $L_{min} < 10^{47}$ erg/s and any lifetime distribution
that provides a large fraction of SHBs at low redshift) predicts
that about $3\%$ of the SHBs observed by BATSE are within $100$ Mpc
while about $1\%$ are within $50$ Mpc. Given that {\it Swift}
detects $\sim 10$ SHBs a year and that its threshold is comparable
to that of BATSE, the expected rate of simultaneous detections of an
SHB and a GW signal is $\sim 0.3 [0.1] ~\rm yr^{-1}$. This rate
depends on the beaming only through the assumption that SHBs are
beamed perpendicular to the binary orbital plane. It could be
further increased by a search for simultaneous GW signals and SHBs
that are detected by {\it Swift} or the Interplanetary Network (IPN)
but are not localized (such bursts are currently not announced and
are not included in the {\it Swift} detection rate that we quote
above).

Simultaneous detection of the inspiral GW signal from a compact
merger and an SHB will provide a conclusive evidence that SHBs
originate from compact mergers and would improve our understanding
of both merger physics and SHBs significantly. GWs can provide a
unique view of the formation, and possibly the operation, of the
inner engine powering burst, difficult to observe via any other
observational method. If GWs from a merger are detected without a
detected SHB, an association may still be secured by the detection
of an on-axis orphan afterglow \citep{Nakar03}. The probability to
detect an off-axis radio orphan afterglow as suggested by
\citet{Levinson02} is low unless the beaming is not significant. The
reason is that the isotropic energy of an SHB that would be detected
by LIGO-I or VIRGO is expected to be $\lesssim 10^{47}$ erg, and if
the beaming is significant, its total energy would be much lower,
and the afterglow is expected to be too dim for detection by the
time that the jet decelerates and spherical emission dominates.

Advanced LIGO (LIGO-II) would be able to detect NS-BH [DNS] mergers
up to a distance of $650[300]$ Mpc \citep{Cutler02}. Thus, if SHBs
originate in such mergers then a robust lower limit for the
detection of GWs by LIGO-II is $20[2] ~\rm yr^{-1}$ and the probable
rate is larger by orders of magnitude. This rate is for detections
of GW signals that are not associated with the prompt gamma-ray
emission from an SHB. As discussed above, simultaneous detections
will increase the LIGO-II range by a factor of 1.5-2.5 to $\approx
1.3[0.6] ~\rm{Gpc}$ \citep{Kochanek93}. So far, {\it HETE-2}
observed one burst at a distance $<700$ Mpc while {\it Swift}
detected at least additional $2$ SHBs at distance $\lesssim 1$ Gpc.
The GBM detector on {\it
GLAST}\footnote{http://f64.nsstc.nasa.gov/gbm/} is expected to have
a threshold that is similar to BATSE and more than half sky field of
view, and thus it is expected to detect at least $5$ SHBs within a
distance of $500$ Mpc every year. The benefit of the simultaneous
operation of LIGO-II and an efficient GRB detector goes beyond the
high likelihood to observe simultaneous SHBs and mergers if they are
associated - it will also be able to disprove the association at
high significance if no simultaneous detection is observed.

SHBs may emit GWs also if the progenitor is not a compact binary
merger. Any progenitor that involves the collapse of a rotating
compact object to a black hole (e.g., the collapse of a rotating
neutron star triggered by accretion; \citealt{MacFadyen05}) will
produce gravitational waves (see, e.g., \citealt{Stark85}). The
amplitude of these waves is highly uncertain. Moreover, the absence
of an accurate signal template will reduce its detectability. Such
GW signals would most likely not be detected even by LIGO-II at
distances much greater than $10$ Mpc (\citealt{Kokkotas05}, and
references therein). If SHBs  result from the collapse of a compact
object that is not triggered by a merger, LIGO-II might detect
several SHBs if they are significantly beamed and/or $L_{min} \ll
10^{49}$ erg/s.

\section{Conclusions}

We have extended previous analysis schemes
\citep{Piran92,Ando04,Guetta05}, combining the observed peak flux
distribution and the  two-dimensional  observed redshift-luminosity
distribution of SHBs, to constrain their progenitor lifetime and
their local rate. We apply this method to the extended sample of
SHBs with redshift constraints presented in \cite{gal-yam05}. Our
conclusions are:

\begin{itemize}
    \item The progenitors of SHBs are dominated by an old
    population. If there is a typical lifetime than it is
    $\tau_*>4[1]$ Gyr at $95\%[99.9\%]$ confidence level.
    If the lifetime distribution is a power-law with index
    $\eta$ then the number of progenitors per
    logarithmic lifetime interval increases significantly with
    lifetime - $\eta>-0.5 [-1]$ at $95\%[99.5\%]$ confidence level.
    Similar results are obtained when only the {\Swift} SHB sample is
    considered. These results were obtained assuming that the luminosity
    function and the lifetime distribution are unimodal.

    \item We derive a lower limit on the local rate of SHBs,
    ${\cal R}_{SHB,obs} \gtrsim 10 ~\rm Gpc^{-3} ~yr^{-1}$.
    This rate is comparable to the BATSE observed (all-sky) rate
    and is a robust lower limit. It is obtained when we use the simplest luminosity
    function that fits the data, $\phi_L \propto L^{-2}$, or
    independently, if we take the {\it Swift} threshold to be comparable to
    that of BATSE, as implied by the SHBs observed by {\it Swift} so far,
    and consider only the {\it Swift} sample.

    \item Assuming that SHB progenitors are the end products of core-collapse SNe, and that
    they are catastrophic, we derive also a rate upper limit - the rate of core-collapse
    SNe at $z \approx 0.7$: $10<{\cal R}_{SHB} < 5 \times 10^5 ~\rm Gpc^{-3}
    yr^{-1}$.

    \item Indications that SHBs are beamed \citep{Fox05} and that
    the luminosity function lower cutoff is $L_{min} \ll 10^{49}$
    erg/s (\S\ref{SEC rate} and \cite{tanvir05}) suggests that the
    local rate is much higher than our lower limit. Considering the
    current best, but highly uncertain, estimates of the beaming and
    $L_{min}$ we get a local rate of ${\cal R}_{SHB} \approx 10^5 ~\rm Gpc^{-3}
    yr^{-1}$.

\end{itemize}

We compare our constraints on the progenitors of SHBs to
observational constraints on DNS mergers. We find that observed DNS
binaries in our galaxy are unlikely to be a representative sample of
SHB progenitors. The reason is that the merger rate of the observed
DNS binaries is dominated by short-lived systems ($\tau \lesssim
100$\,Myr) while SHB progenitors are much older ($\tau\gtrsim
4$\,Gyr). We are not aware of any obvious selection effect that
would prevent the detection of recently formed yet long-lived DNS
binaries and therefore we conclude that our results disfavor DNS
systems as the progenitors of SHBs. However, given the limited
understanding of the formation of DNS systems and the possibility
that some of the formation channels result in binaries that cannot
be detected, we cannot exclude the possibility that DNS systems end
their life as SHBs. If they do, then we expect a large number of
undetected old DNS systems in our galaxy. We consider also the
theoretical constraints on BH-NS mergers and find that there are no
robust constraints that prevent these from being the progenitors of
SHBs (based on the lifetime and the rate), due to the uncertain
properties of the BH-NS population.

If SHBs originate from either BH-NS or DNS mergers, then the
prospects for detecting GWs by first generation detectors and in the
future planned advance LIGO (LIGO-II) are promising:
\begin{itemize}
    \item Our robust lower limit on the local rate of SHBs implies
    that LIGO-II will detect at least $20$ merging BH and NS pairs
    with a mass ratio of a few, per year, if they are the progenitors of SHBs.
    A minimal rate of  $2$ DNS (or BH-NS with similar masses) mergers per
    year is predicted if these are the sources of SHBs. If an efficient GRB detector
    (e.g. GLAST) will be  operational contemporaneously with LIGO-II,
    numerous simultaneous detections are expected. The absence of such
    detections will disprove models of SHBs are as
    compact binary mergers.

    \item The highest possible rate of SHBs that is consistent
    with the upper limit on the merger rate of compact binaries ($10^4 ~\rm Gyr^{-3} ~yr^{-1}$)
    predicts a LIGO-I and VIRGO detection rate
    of $\sim 3 ~\rm yr^{-1}$ if  SHBs are mergers of BH-NS with a mass ratio of a few,
    and $\sim 0.3 ~\rm yr^{-1}$ if they are DNS mergers or mergers of BH-NS with similar masses.
    This is the rate for GW detection only, with no associated SHB detection by
    {\it Swift}.

    \item  The predicted probability for a simultaneous detection of GWs by LIGO-I or VIRGO and
    an SHB by {\it Swift} in one year of contemporary operation might not be negligible:
    $\sim 30\%$ for a BH-NS merger and $\sim 10\%$ for a DNS
    merger. This probability is derived assuming $\phi_L \propto
    L^{-2}$ with a lower luminosity cutoff $L_{min} = 10^{47}$ erg/s.

    \item The probability for a simultaneous detection can be
    increased if a search for GW signals will be carried out at
    times that SHBs are detected, but not localized, by {\it Swift} or by
    any of the IPN spacecraft. If SHBs are not strongly beamed,
    this may significantly increase the probability of LIGO-I to
    detect a single event.
\end{itemize}

%%%%%%%%%%%%%%%%%%%%%%%%%%%%%%%%%%%%%%%%%%%%%%%%%%%%%%%%%%%%%%%%%%%%%%

\section*{Acknowledgments}

We are grateful to E. S. Phinney and E. Ofek for illuminating
discussions and advice. We thank D. Frail, S. Kulkarni, R. Sari., B.
Cameron, M. Milosavljevi\'c, H. Pfeiffer, D. Palmer, T. Piran, S.
Ando and C. Miller for helpful discussions. E.N. was supported by a
senior research fellowship from the Sherman Fairchild Foundation.
A.G. acknowledges support by NASA through Hubble Fellowship grant
\#HST-HF-01158.01-A awarded by STScI, which is operated by AURA,
Inc., for NASA, under contract NAS 5-26555.

%%%%%%%%%%%%%%%%%%%%%%%%%%%%%%%%%%%%%%%%%%%%%%%%%%%%%%%%%%%%%%%%%%%%%%
%\bibliographystyle{apj}
%\bibliography{ms}

%%%%%%%%%%%%%%%%%%%%%%%%%%%%%%%%%%%%%%%%%%%%%%%%%%%%%%%%%%%%%%%%%%%%%%
%%%%%%%%%%%%%%%%%%%%%%%%%%%%%%%%%%%%%%%%%%%%%%%%%%%%%%%%%%%%%%%%%%%%%%
%%%%%%%%%%%%%%%%%%%%%%%%%%%%%%%%%%%%%%%%%%%%%%%%%%%%%%%%%%%%%%%%%%%%%%

\end{document}